\documentclass[aps,prd,floatfix,nofootinbib]{revtex4} 
\usepackage{graphics}
\usepackage{graphicx}

\usepackage{latexsym}
\usepackage[cp866]{inputenc}
\usepackage[english]{babel}
\input epsf

\begin{document}

\title{\boldmath Electroweak production of $\chi_{Q1}$ states
in $e^+e^-$ collisions: A brief review}
\author{N.~N. Achasov,$^{1,}$\footnote{achasov@math.nsc.ru}
A.~V. Kiselev,$^{1,2,}$\footnote{kiselev@math.nsc.ru} and G.~N.
Shestakov,$^{1,}$\footnote{shestako@math.nsc.ru}}
\affiliation{$^1$\,Laboratory of Theoretical Physics, S.~L. Sobolev
Institute for Mathematics, 630090, Novosibirsk, Russia \\
$^2$\,Novosibirsk State University, Novosibirsk 630090, Russia}


\begin{abstract}
A brief review of the available experimental and theoretical results
on the production of the $\chi_{Q1}$ states in $e^+e^-$ annihilation
and photon-photon $\gamma\gamma^*$ interactions is presented. Future
data on the production of the $\chi_{c1}(1P)$, $\chi_{c1}(3872)$,
$\chi_{c1} (4140)$, $\chi_{c1}(4274)$, $\chi_{c1}(4685)$, $\chi_{b1}
(1P)$, $\chi_{b1}(2P )$, and $\chi_{b1}(3P)$ resonances in $e^+e^-$
annihilation and $\gamma\gamma^*$ interactions will help the
development and unification of theoretical predictions related to
the electroweak decays of heavy quarkonia, the reduction of their
spread and model uncertainties.
\end{abstract}

\maketitle

\section{Introduction}

In the report made at the $PhiPsi\ {\it 2019}$ conference in
Novosibirsk \cite{Ac19} (see  also references therein), a point of
view was presented on the nature of the $X(3872)$ state \cite{Cho03}
as the first excited state in the axial-vector charmonium family,
$X(3872)=\chi_{c1}(3872)=\chi_{c1}(2P)$, and discussed as probes to
study the internal structure of the $\chi_{c1}$ and $\chi_{b1}$
states electroweak reactions $e^+e^-\to\chi_{c1}/\chi_{b1}$.
Recently, important progress has been made in this direction. The
BESIII and Belle collaborations performed the first measurements of
the direct production of the $\chi_{c1}(1P)$ in $e^+e^- $
annihilation \cite{Ab22} and the $X(3872)$ formation in
photon-photon interactions \cite{Ter21}.
In addition, the BESIII collaboration \cite{Ab22a} has improved the
upper limit for the electronic width of the $X(3872)$ resonance by a
factor of about 13 compared to the previous limit. There is a huge
theoretical discussion about the nature of the $X(3872)$ in the
literature, for a review see Refs. \cite{Bu13,KR15,AR15,Le17,Es17,
Gu18,Br20,Ci22,Co22} and references therein. Various scenarios are
tested for it, from the resonance $c\bar c$ state to a loosely bound
$D\bar D^*$ molecule. However, understanding the nature of the
$X(3872)$ is still an open challenge. Let us take a few interesting
statements as an example. From Ref. \cite{Le17}: ``Although the
technology for describing the $X(3872)$ as a primarily $D^0\bar
D^{*0}$ molecule is quite mature, solid reasons exist for
questioning this interpretation.'' From Ref. \cite{Br20}: ``In this
review we only state that at present there is no full understanding
of the production rates of shallow molecular states. Implications of
the molecular scenario are contrasted with different quark model
approaches in Ref. [540].'' From Ref. \cite{Ci22}: ``Reasonable
cross sections can be obtained in either $c\bar c$ or molecular
$D\bar D^*$ scenarios for $X(3872)$. Also a hybrid scenario is not
excluded.'' The situation is developing dynamically and each new
evidence about the nature of the $X(3872)$ state is highly
important.

This work gives a brief overview of the current situation related to
the electroweak production of $\chi_{Q1}$ states in $e^+e^-$
collisions. Section II presents the available experimental data.
Theoretical estimates for the widths of the $\chi_{Q1}\to e^+e^-$
and $\chi_{c1}\to\gamma\gamma^*$ decays are discussed in Secs. III
and IV, respectively. The conclusions are presented in Sec. V.

\section{Experimental data}

The first estimates of the decay width into $e^+e^-$ for the $P
$-wave states of heavy quarkonia $\chi_{Q1}$ ($Q=c,b$) were obtained
as far back as the late 1970s \cite{KK78,KKS79,Kuh81}. It was
assumed that the direct creation of resonances with $J^{PC}=1^{++}$
in $e^+e^-$ annihilation can occur due to two different mechanisms:
via $Z^0$ exchange, $ e^+e^-\to Z^0\to \chi_{Q1}$, and through two
virtual photons, $e^+e^-\to\gamma^*\gamma^*\to\chi_{Q1}$.
Calculations of the electronic decay width, $\Gamma(\chi_{Q1}\to
e^+e^-)$, carried out in Refs. \cite{KK78,KKS79,Kuh81,Re82,YZ12,
DGHN14,KV16,CKT16,CKT17} have demonstrated a significant dependence
of the result on the assumptions made and on the choice of
parameters. For example, for $\Gamma(\chi_{c1} (1P)\to e^+e^-)$
(i.e., for electronic width of the ground axial-vector state of
charmonium) the values were predicted in the range from 0.044 eV to
0.41 eV \cite{KKS79,YZ12,DGHN14,KV16,CKT16}, and for the excited
state $\chi_{c1}(3872)$ it was found that $\Gamma(\chi_{c1}(3872)\to
e^+e^-)\gtrsim 0.03$ eV \cite{DGHN14}. More details about the
theoretical estimates of $\Gamma(\chi_{Q1}\to e^+e^-)$ will be
discussed in Sec. III.

An upper limit for the electronic width of the $\chi_{c1}(3872)$ on
the $\mathcal{O}$(eV) level (namely, $<4.3$ eV) was first obtained
in 2015 by the BESIII collaboration \cite{Ab15} while studying the
reaction $e^+e^-\to \gamma_{ \mbox{\scriptsize{ISR}}}\chi_{c1}(3872
)\to\gamma_{\mbox{\scriptsize{ISR}}}\pi^+\pi^-J/\psi$ with
$\gamma_{\mbox{\scriptsize{ISR}}}$ photon emission from the initial
state.
More recently, the BESIII collaboration lowered an upper limit on
$\Gamma(\chi_{c1}(3872)\to e^+e^-)$ up to 0.32 eV \cite{Ab22a}.
As for the ground state $\chi_{c1}(1P)$, its direct production in
$e^+e^-$ annihilation with a significance of $5.1\sigma$ was first
reported by the BESIII collaboration most recently \cite{Ab22}. For
the electron width of the $\chi_{c1}(1P)$ the following value was
obtained \cite{Ab22}:
\begin{eqnarray}\label{Eq1} \Gamma(\chi_{c1}(1P)\to e^+e^-)=\left(
0.12^{+0.13}_{-0.08}\right)\ \mbox{eV}.\end{eqnarray} It was the
result of processing data on the interference between the signal
process $e^+e^-\to\chi_{c1}(1P)\to\gamma J/\psi\to\gamma \mu^+\mu^-$
and coherent background processes $e^+e^-\to\gamma_{
\mbox{\scriptsize{ISR}}}J/\psi\to\gamma_{\mbox{\scriptsize{ISR}}}
\mu^+\mu^-$ and $e^+e^-\to\gamma_{\mbox{\scriptsize{ISR}}}\mu^+
\mu^-$ \cite{Ab22,CKT16,CKT17}.

As is known, for the axial-vector state $1^{++}$ there is a
selection rule \cite{Lan48,Yan50} that excludes its decay into two
real photons. However, such a state can be produced in two-photon
collisions, when one or both photons are virtual (virtual photon
$\gamma^*\equiv \gamma^*(Q^2)$, where $-Q^2$ is the square of its
invariant mass). The result of the first experiment recently
performed by the Belle collaboration \cite{Ter21} to search for
$\chi_{c1}(3872)$ production in $\gamma\gamma^*$ collisions is the
following estimate based on the observation of three
$\chi_{c1}(3872)$ candidates with a significance of $3.2\sigma$ and
assuming the $Q^2$ dependence predicted by a $c\bar c$ meson model:
\begin{eqnarray}\label{Eq2}\widetilde{\Gamma}(\chi_{c1}(3872)\to
\gamma\gamma)\mathcal{B}(\chi_{c1}(3872)\to\pi^+\pi^- J/\psi)=
\left(5.5 \pm^{4.1}_{3.8}(\mbox{stat})\pm0.7\,(\mbox{syst})\right)\
\mbox{eV}.\end{eqnarray} Here $\widetilde{\Gamma}(\chi_{c1} (3872)
\to\gamma\gamma)$ is so-called the reduced $\gamma\gamma$ decay
width defined (with use of the short notation $\widetilde{\Gamma}_{
\gamma\gamma}$) as \cite{Re84,Ols87,Ca87, Aih88,SBG98,Ter21}
\begin{eqnarray}\label{Eq3}\widetilde{\Gamma}_{\gamma\gamma}\equiv
\lim_{Q^2\to0}\frac{M^2}{Q^2}\Gamma^{TL}_{\gamma\gamma^*}(Q^2),
\end{eqnarray} where $M$ is the mass of the resonance and $\Gamma^{T
L}_{\gamma\gamma^*}(Q^2)$ is the $\gamma\gamma^*$ decay width
corresponding to a formation of the resonance  by a transverse
(real) photon $\gamma$ and a longitudinal (virtual) photon
$\gamma^*$. For the reduced width $\widetilde{\Gamma}(\chi_{c1}
(3872)\to\gamma\gamma)$ in Ref. \cite{Ter21} a possible range of
values from 20 to 500 eV is indicated. If one utilizes the updated
branching ratio $\mathcal{B}(\chi_{c1}(3872)\to\pi^+\pi^-
J/\psi)=(3.8 \pm1.2)\%$ \cite{PDG22}, then this interval, while
remaining wide, moves slightly towards larger values:
\begin{eqnarray}\label{Eq4}24\ \mbox{eV}<\widetilde{\Gamma}(\chi_{
c1}(3872)\to\gamma\gamma)<615\ \mbox{eV}.\end{eqnarray}
Investigations of the $\chi_{c1}(3872)$ production in $\gamma
\gamma^*$ collisions will be continued at higher luminosities at the
Belle II facility \cite{Ter21}. Theoretical estimates of the width
$\widetilde{\Gamma}(\chi_{c1}(3872)\to\gamma\gamma)$ we will discuss
in Sec. IV.

\section{\boldmath Direct production of $\chi_{Q1}$ states in
$e^+e^-$ annihilation}

The width of the decay $\chi_{Q1}\to Z^0\to e^+e^-$ due to the weak
neutral current ($Z^0$ exchange) at $m^2_{\chi_{Q1}}/m^2_{Z^0}\ll1$
and $m_e=0$ has the form \cite{KK78,Kuh81,Re82,YZ12,CKT16}
\begin{eqnarray}\label{Eq5}
\Gamma(\chi_{Q1}\to Z^0\to e^+e^-)=\frac{3G^2_F}{4\pi^2}|R'_{
\chi_{Q1}}(0)|^2(g^{e2}_a+g^{e2}_v),
\end{eqnarray}
where the Fermi constant $G_F=1.116\times10^{-5}$ GeV$^{-2}$,
$g^e_a=1$, $g^e_v=-1+4 \sin^2\theta_W$, and $R'_{\chi_{Q1}}(0)$ is
the derivative of the $\chi_{Q1}$ radial wave function at the origin
[of course, this function is the same for all states
$\chi_{QJ=0,1,2}(nP)$ at given $n$ ($n=1,2,...$)]. For estimates, we
put $\sin^2\theta_W=1/4$, and also $|R'_{\chi_{ c1}}(0)|^2\approx
0.1$ GeV$^{5}$ for the $\chi_{c1}(1P)$ state and $|R'_{\chi_{b1}}(0)
|^2\approx2$ GeV$^{5}$ for the more compact $\chi_{b1}(1P)$ state
(tabulated values of the quarkonium radial wave functions at the
origin can be found, for example, in Refs. \cite{EQ95,EQ19}). From
this we have
\begin{eqnarray}\label{Eq6} \Gamma(\chi_{c1}(1P)\to Z^0\to
e^+e^-)\approx10^{-3}\ \mbox{eV}\,\ \ \mbox{and}\,\ \
\Gamma(\chi_{b1}(1P)\to Z^0\to e^+e^-)\approx2\times 10^{-2}\
\mbox{eV}.\end{eqnarray}

As already noted in Sec. II, the direct production of an
axial-vector resonance with $J^{PC}=1^{++}$ in $e^+e^-$ annihilation
can occur due to two different mechanisms: via $Z^0$ exchange [see
Eqs. (\ref{Eq5}) and (\ref{Eq6})] and through two virtual photons.
The decay width $\chi_{Q1}\to e^+e^-$ corresponding to these two
mechanisms and their interference can be represented in the
so-called logarithmic approximation in the following form (details
of calculations, in particular, in the vector dominance model,
discussions about the meaning and method of regularizing the
logarithmic singularity, as well as a number of modifications of the
simple formula below, taken from \cite{YZ12}, can be found in Refs.
\cite{KKS79,Kuh81,Re82,YZ12,DGHN14,KV16,CKT16,CKT17,BGR76,Jac77,
Nov78}):
\begin{eqnarray}\label{Eq7}\Gamma(\chi_{Q1}\to
e^+e^-)=\frac{3}{4\pi^2}|R'_{\chi_{Q1}}(0)|^2\left[G^2_F(g^{e2}_a+
g^{e2}_v)\mp\frac{2\sqrt{2}e^2_Q\alpha^2G_F}{m^2_Q}g^e_a\mbox{Re}
[f_1]+\frac{2e^4_Q\alpha^4}{m^4_Q}|f_1|^2\right].
\end{eqnarray}
Here ``$-$'' corresponds to $Q=c$ and ``$+$'' to $Q=b$; $e_c=2/3$
and $e_b=-1/3$; the coefficient $f_1=4\ln(m_Q/\omega)$. The
parameter $\omega$ has no unambiguous interpretation. On an
intuitive level, $\omega$ is defined, for example, as the binding
energy of $\chi_{Q1}$, $\omega=2m_Q-M_{\chi_{Q1}}$
\cite{KKS79,BGR76,Jac77,CKT17}, or as the binding energy, but with
the opposite sign $\omega=M_{\chi_{Q1}}-2m_Q$ \cite{YZ12} [in both
cases, the appearance in the decay amplitude  of the imaginary part
due to $\ln(m_Q/\omega)$ at $\omega<0$ is rather obscure], or one
supposes that $\omega\sim1/R\simeq(300$--$500)$ MeV, where $R$ is
the quarkonium radius \cite{Nov78}, or $\omega\simeq(300$--$500)$
MeV is considered to be the characteristic virtuality scale of one
of the intermediate virtual photons, i.e., of the soft virtual
photon in the decay $\chi_{Q1}\to\gamma^*\gamma^*\to e^+e^-$
\cite{KV16}. Here we do not dwell on the consideration of
contributions from intermediate states like $\gamma^*J/\psi$,
$\gamma^*\psi(2S)$, $\gamma^*\Upsilon(1S)$, etc. These contributions
are discussed in detail in Refs.
\cite{KKS79,Kuh81,YZ12,DGHN14,KV16,CKT16,CKT17}, where various
values are predicted for them (in this connection see also Ref.
\cite{FN1}). In order to have concrete numerical estimates before
our eyes, we set, following \cite{YZ12},
$\omega=M_{\chi_{Q1}}-2m_Q$, $M_{\chi_{c1}(1P )}=3.51$ GeV,
$m_c=1.65$ GeV, $M_{\chi_{b1}(1P)}=9.89$ GeV, and $m_b=4.67$ GeV.
Then according to Eq. (\ref{Eq7}) we have
\begin{eqnarray}\label{Eq8}
\Gamma(\chi_{c1}(1P)\to e^+e^-)\approx(0.00095-0.01720+0.07817)\
\mbox{eV}\approx0.06192\ \mbox{eV},
\end{eqnarray}
\begin{eqnarray}\label{Eq9}
\Gamma(\chi_{b1}(1P)\to e^+e^-)\approx(0.0189+0.0111+0.0016)\
\mbox{eV}\approx0.0316\ \mbox{eV},
\end{eqnarray}
where in parentheses are the contributions of individual terms. The
estimate in Eq. (\ref{Eq8}) does not contradict the BESIII data,
taking into account their errors, see Eq. (\ref{Eq1}). Although the
electron widths in Eqs. (\ref{Eq8}) and (\ref{Eq9}) formally turn
out to be of the same order, the mechanism of their formation upon
going from $\chi_{c1}(1P)$ to $\chi_{b1}(1P)$ is undergoing a
fundamental changing. Really, the contribution of the $\chi_{
b1}(1P)\to\gamma^* \gamma^*\to e^+e^-$ mechanism [see the third term
in Eq. (\ref{Eq9})] decreases compared to the contribution of the
$\chi_{c1}(1P)\to\gamma^*\gamma^*\to e^+ e^-$ mechanism [see the
third term in Eq. (\ref{Eq8})] by about 50 times owing to the factor
$|R'_{\chi_{Q1} }(0)|^2(e_Q/m_Q)^4$ [in so doing the factor
$(e_Q/m_Q)^4$ decreases by about a thousand times]. Thus, it is
natural to expect that the contribution of the weak neutral current
dominates in the decay $\chi_{b1}(1P)\to e^+e^-$, while the decay
$\chi_{c1}(1P)\to e^+e^-$ is dominated by the two-photon transition.

What can be at least qualitatively said about the electronic widths
of excited states $\chi_{c1}(3872)$, $\chi_{c1}(4140)$,
$\chi_{c1}(4274)$, $\chi_{c1}(4685)$, and $\chi_{b1}(2P)$,
$\chi_{b1}(3P)$ \cite{PDG22}? For all states $\chi_{b1}(nP)$
($n=1,2,...$) it is reasonable to assume that their electronic
widths are determined mainly by the $Z^0$ exchange mechanism and
therefore must be controlled by the corresponding values of
$|R'_{\chi_{b1}(nP)}(0)|^2$. Calculations in potential models
\cite{EQ95,EQ19} show that $|R'_{\chi_{b1}(nP)}(0)|^2$ either grows
weakly with $n$ or remains virtually constant. Therefore, the widths
$\Gamma(\chi_{b1}(nP)\to e^+e^-)$ should be expected to be
approximately the same. For the states $\chi_{c1}(nP)$, $|R'_{\chi_{
c1}(nP)}(0)|^2$ behave with $n$ increasing by a similar way
\cite{EQ95,EQ19}. However, the dominant two-photon transition
mechanism $\chi_{c1}(nP)\to\gamma^* \gamma^*\to e^+e^-$ can
significantly depend on $n$ due to the different dependence on $n$
of the contributions of the intermediate states $\gamma^*J/\psi$,
$\gamma^*\psi(2S)$, etc. The measurement of the widths
$\Gamma(\chi_{Q1}(nP)\to e^+e^-)$ is a very difficult task [see the
description of the BESIII experiment on measuring
$\Gamma(\chi_{c1}(1P)\to e^+e^-)$ \cite{Ab22}]. But there is no
doubt that each such measurement is an important step towards
understanding internal structure of heavy quarkonia.

The total cross section for the $\chi_{Q1}$ resonance production
with unpolarized $e^+e^-$ beams is given by
\begin{eqnarray}\label{Eq9b}
\sigma(e^+e^-\to\chi_{Q1};E)=\frac{3\pi}{E^2}\frac{\Gamma(\chi_{Q1}\to
e^+e^-)\,\Gamma_{\chi_{Q1}}}{(M_{\chi_{Q1}}-E)^2+\Gamma^2_{\chi_{Q1}}/4},
\end{eqnarray}
where $E$ is the energy in the $e^+e^-$ center-of-mass system and
$\Gamma_{\chi_{Q1}}$ is the total width of the $\chi_{Q1}$ state. If
we put as an example $\Gamma(\chi_{Q1}\to e^+e^-)=0.1$ eV and
$\Gamma_{\chi_{Q1}}=1$ MeV, then for the $R_{\chi_{Q1}}$ value in
the peak of the $\chi_{Q1}$ resonance we find
\begin{eqnarray}\label{Eq9c}
R_{\chi_{Q1}}=\frac{\sigma(e^+e^-\to\chi_{Q1};E)}{\sigma(e^+e^-\to
\gamma^*\to\mu^+\mu^-;E)}=\frac{9}{\alpha^2}\frac{\Gamma(\chi_{Q1}\to
e^+e^-)}{\Gamma_{\chi_{Q1}}}\approx0.017.\end{eqnarray}

Presently upgrading the SuperKEKB electron-positron collider with
polarized electron beams is planned that opens a new physics program
owing to precision neutral current measurements \cite{As22}; see
also Ref. \cite{Ya22}. Dependence of the amplitude of $\chi_{Q1}$
production, $\mathcal{M}(e^+e^-\to \chi_{Q1})$, on the sign of the
electron helicity $\lambda$ (or on the direction of the polarization
of the electron beam) is determined by  the vector part of the weak
neutral current. The corresponding contribution is proportional to
$g^e_v=-1+4\sin^2 \theta_W$, i.e., deviation of $4\sin^2\theta_W$
from 1. Setting $\sin^2\theta_W=0.231$ [32], we get
\begin{eqnarray}\label{Eq9a}
\mathcal{M}_{\lambda=\pm1}(e^+e^-\to\chi_{Q1})\propto[(\pm
g^e_v+g^e_a)A^{\chi_{Q1}}_{Z^0}+A^{\chi_{Q1}}_{\gamma^*\gamma^*}]
=[(\mp0.076+1)A^{\chi_{Q1}}_{Z^0}+A^{\chi_{Q1}}_{\gamma^* \gamma^*}
],\end{eqnarray}  where $A^{\chi_{Q1}}_{\gamma^*\gamma^*}$ and
$A^{\chi_{Q1}}_{Z^0}$ are the amplitudes of the $\chi_{Q1}\to
e^+e^-$ transitions via $\gamma^*\gamma^*$ and $Z^0$ mechanisms
respectively. If the contribution of the amplitude $A^{\chi_{Q1}}_{
\gamma^*\gamma^*}$ dominates, then the effect associated with the
polarization is very small. If the contribution of the amplitude
$A^{\chi_{Q1}}_{Z^0}$ dominates, which is very likely for the
$\chi_{b1}$ states, then due to the interference phenomena, the
relative effect of polarization (the effect of parity violation) can
be up to 15\%.


\section{\boldmath Direct annihilation transition $\chi_{c1}\to\gamma\gamma^*$}

According to Ref. \cite{Ca87}, the decay width of the $^3P_1$
nonrelativistic bound state of charmonium with mass $M$ into
$\gamma\gamma^*(Q^2)$ for small $Q^2$ can be represented as
\begin{eqnarray}\label{Eq10}
\Gamma(^3P_1\to\gamma\gamma^*(Q^2))=192 \alpha^2e^4_c\frac{|R'
(0)|^2}{M^4}\frac{Q^2}{M^2}\equiv \widetilde{\Gamma}(^3P_1\to\gamma
\gamma)\frac{Q^2}{M^2}, \end{eqnarray} where
$\widetilde{\Gamma}(^3P_1\to\gamma\gamma)$ is the so-called reduced
$\gamma\gamma$ decay width \cite{Re84,Ols87,Ca87,Aih88, SBG98,
Ter21}, see Eq. (\ref{Eq4}), and $R'(0)$ is the derivative of the
radial wave function at the origin for the corresponding $P$-wave
state of charmonium. The decay widths of the $^3P_0$ and $^3P_2$
states into $\gamma \gamma$ have the form \cite{BGK76}
\begin{eqnarray}\label{Eq11}
\Gamma(^3P_0\to\gamma\gamma)=432 \alpha^2e^4_c\frac{|R'(0)|^2}{M^4},
\end{eqnarray}
\begin{eqnarray}\label{Eq12}
\Gamma(^3P_2\to\gamma\gamma)=\frac{576}{5}
\alpha^2e^4_c\frac{|R'(0)|^2}{M^4}.
\end{eqnarray}
A discussion of $\alpha_s$ corrections to these relations can be
found, for example, in Refs. \cite{Eck08,Zho21}. The value of the
ratio \cite{BGK76}
\begin{eqnarray}\label{Eq13}
\frac{\Gamma(^3P_2\to\gamma\gamma)}
{\Gamma(^3P_0\to\gamma\gamma)}=\frac{4}{15}\simeq0.27
\end{eqnarray}
for the case of the states $\chi_{c2}(1P)$ and $\chi_{c0}(1P)$ is in
good agreement with the available data \cite{Eck08,Ab17,PDG22}. From
Eqs. (\ref{Eq10}) and (\ref{Eq12}) we have
\begin{eqnarray}\label{Eq14}
\frac{\widetilde{\Gamma}(^3P_1\to\gamma\gamma)
}{\Gamma(^3P_2\to\gamma\gamma)}=\frac{5}{3}.
\end{eqnarray}
According the Particle Data Group \cite{PDG22}
$\Gamma(\chi_{c2}(1P)\to\gamma\gamma)\simeq0.56$ keV. Hence for
$\widetilde{\Gamma}(\chi_{c1}(1P)\to\gamma\gamma)$ we obtain the
estimate
\begin{eqnarray}\label{Eq15}
\widetilde{\Gamma}(\chi_{c1}(1P)\to\gamma\gamma)
\approx\frac{5}{3}\times0.56\ \mbox{keV}\approx0.93\ \mbox{keV}.
\end{eqnarray}
For the excited $2^3P_2$ state $\chi_{c2}(3930)$ it is known
\cite{PDG22} that
\begin{eqnarray}\label{Eq16}
\Gamma(\chi_{c2}(3930)\to\gamma\gamma)
\mathcal{B}(\chi_{c2}(3930)\to D\bar D)=(0.21\pm0.04)\ \mbox{keV}.
\end{eqnarray}
If $\mathcal{B}(\chi_{c2}(2P)\to D\bar D)\approx1$ [see however
Refs. \cite{AKa17,Wa22}; future experiments will help refine the
value of $\mathcal{B}(\chi_{c2}(2P)\to D\bar D)$], then for its
$2^3P_1$ partner $\chi_{c1}(3872)$ from Eqs. (\ref {Eq14}) and
(\ref{Eq16}) we obtain the following estimate for the corresponding
reduced $\gamma\gamma$ decay width:
\begin{eqnarray}\label{Eq17}
\widetilde{\Gamma}(\chi_{c1}(3872)
\to\gamma\gamma)\approx\frac{5}{3}\times0.21\ \mbox{keV}\approx350\
\mbox{eV}.
\end{eqnarray}
This estimate falls within the range of possible values for
$\widetilde{\Gamma}(\chi_{c1}(3872)\to\gamma\gamma)$ specified in
Eq. (\ref{Eq4}) based on the data from Belle \cite{Ter21}.

Recently,  in the light-front approach, the following values for the
reduced two-photon widths of the $\chi_{c1}(1P)$ and $\chi_{c1}(3872
)$ states have been predicted \cite{LLV22}: $\widetilde{\Gamma}(
\chi_{c1}(1P)\to\gamma\gamma) =(3\pm0.5)$ keV and $\widetilde{
\Gamma}(\chi_{c1}(3872)\to\gamma \gamma)=(3\pm1 )$keV. Both these
values significantly exceed the above theoretical and experimental
estimates.

The $\chi_{c1}(3872)$ state has a finite width $\approx1$ MeV, which
is due to strong and radiative decays, among which the dominant
decay is $\chi_{c1}(3872)\to(D^{*0}\bar D^0+\bar D ^{*0} D^0)$
\cite{PDG22}. Therefore, the estimates of the contributions to the
width $\widetilde{\Gamma}(\chi_{c1}(3872)\to\gamma\gamma)$
corresponding to rescattering mechanisms like $\chi_{c1}(3872)
\to(D^*\bar D+\bar D^*D)\to\gamma \gamma^*$ are of quite natural
interest. So far, there are no such estimates.

\section{Conclusion}

The recent data on the $\chi_{c1}(1P)\to e^+e^-$ and $\chi_{c1}
(3872)\to\gamma\gamma^*$ decays are related to subtle questions of
the electroweak interactions of heavy quarkonia. Their refinement
and extension to other $\chi_{Q1}$ states will certainly lead to a
new wave of theoretical research and predictions within the
framework of various potential nonrelativistic QCD models currently
in use, as well as to the development of new ideas and methods of
studying the nature of heavy quarkonia. The restrictions imposed by
experiment will serve as an effective tool for selecting theoretical
models.

The BESIII experiment \cite{Ab22} demonstrates that with the current
generation of electron-positron colliders, the observation of the
direct production of the $\chi_{c1}$ in $e^+e^-$ annihilation is
possible with the use of the interference phenomena. It is to be
hoped that this method will also be successfully used in subsequent
measurements of the $\chi_{Q1}\to e^+e^-$ decay widths at the BESIII
and Belle II installations and at the future Super Charm-Tau factory
\cite{Bo13}. \vspace{0.2cm}

\begin{center} {\bf ACKNOWLEDGMENTS} \end{center}

The work was carried out within the framework of the state contract
of the Sobolev Institute of Mathematics, Project No. FWNF-2022-0021.




\begin{thebibliography}{99}
\bibitem{FN1}  The predictions \cite{KK78,KKS79,Kuh81,Re82,YZ12,
DGHN14,KV16,CKT16,CKT17} for the widths of the $\chi_{cJ}\to e^+e^-$
decays caused by the contributions from the $\gamma^* \gamma^*$,
$\gamma^*J/\psi$, and $\gamma^*\psi'$ intermediate states vary
considerably from each other. For the $\chi_{c1}(1P)$ state, they
differ approximately by a factor of 5, and for the $\chi_{c2}(1P)$,
the discrepancy reaches almost 2 orders of magnitude. This issue has
not yet been resolved. In those cases where the estimates turn out
to be generally consistent with each other in order of magnitude
\cite{YZ12, DGHN14,KV16}, the difference in the approaches used in
this case and differences in the estimates of individual
contributions still do not allow us to prefer one of the
predictions. In such a situation, we utilize a more or less definite
formula (\ref{Eq7}) to represent specific numbers.
\end{thebibliography}
\end{document}